\documentclass[algorithms,article,accept,moreauthors,pdftex,10pt,a4paper]{Definitions/mdpi} 

\usepackage[utf8]{inputenc}
\usepackage{booktabs} % For formal tables
\usepackage{lipsum} 
\usepackage{balance}
\usepackage{paralist}
\usepackage{filecontents}
\usepackage{etoolbox}
\usepackage{mathrsfs}
\usepackage{bm}
\usepackage{setspace}
\usepackage{amssymb}
\usepackage{graphicx}
\usepackage{caption}
\usepackage{subcaption}
\usepackage[ruled]{algorithm2e}
\usepackage{algpseudocode}
\usepackage{soul}

\setitemize{parsep=6pt,itemsep=0pt,leftmargin=*,labelsep=5.5mm}
\setenumerate{parsep=6pt,itemsep=0pt,leftmargin=*,labelsep=5.5mm}
\setlist[description]{itemsep=0mm}
\firstpage{1} 
\pubvolume{11}
\issuenum{9}
\articlenumber{137}
\pubyear{2018}
\copyrightyear{2018}
\history{Received: 8 August 2018; Accepted: 10 September 2018; Published: 13 September 2018}
\Title{Learning Heterogeneous Knowledge Base Embeddings for Explainable Recommendation}

\Author{Qingyao Ai $^{1}$,%Please carefully check the accuracy of names and affiliations. Changes will not be possible after proofreading. 
 Vahid Azizi $^{2}$, Xu Chen$^{3}$, and Yongfeng Zhang $^{2}$}

\AuthorNames{Qingyao Ai, Vahid Azizi, Xu Chen, Yongfeng Zhang}

\address{%
$^{1}$ \quad College of Information and Computer Sciences, University of Massachusetts Amherst, Amherst, MA 01002, USA; aiqy@cs.umass.edu\\
$^{2}$ \quad Department of Computer Science, Rutgers University, 110 Frelinghuysen Rd, Piscataway Township, 08854 NJ, USA; va190@cs.rutgers.edu, yongfeng.zhang@rutgers.edu\\
$^{3}$ \quad School of Software, Tsinghua University, Beijing 100084, China; xu-ch14@mails.tsinghua.edu.cn}

\abstract{Providing model-generated explanations in recommender systems is important to user experience. State-of-the-art recommendation algorithms --- especially the collaborative filtering (CF)-based approaches with shallow or deep models --- usually work with various unstructured information sources for recommendation, such as textual reviews, visual images, and various implicit or explicit feedbacks. Though structured knowledge bases were considered in content-based approaches, they have been largely ignored recently due to the availability of vast amounts of data and the learning power of many complex models. However, structured knowledge bases exhibit unique advantages in personalized recommendation systems. When the explicit knowledge about users and items is considered for recommendation, the system could provide highly customized recommendations based on users' historical behaviors and the knowledge is helpful for providing informed explanations regarding the recommended items. A great challenge for using knowledge bases for recommendation is how to integrate large-scale structured and unstructured data, while taking advantage of collaborative filtering for highly accurate performance. Recent achievements in knowledge-base embedding (KBE) sheds light on this problem, which makes it possible to learn user and item representations while preserving the structure of their relationship with external knowledge for explanation. In this work, we propose to explain knowledge-base embeddings for explainable recommendation. Specifically, we propose a knowledge-base representation learning framework to embed heterogeneous entities for recommendation, and based on the embedded knowledge base, a soft matching algorithm is proposed to generate personalized explanations for the recommended items. Experimental results on real-world e-commerce datasets verified the superior recommendation performance and the explainability power of our approach compared with state-of-the-art baselines.}


\begin{document}
%%%%%%%%%%%%%%%%%%%%%%%%%%%%%%%%%%%%%%%%%%
%% Only for the journal Gels: Please place the Experimental Section after the Conclusions

%%%%%%%%%%%%%%%%%%%%%%%%%%%%%%%%%%%%%%%%%%
%!TEX root=paper.tex

\section{Introduction}
Most of the existing collaborative filtering %Is the capital neccesary? PLease confirm the whole paper.
(CF)-based recommendation systems work with various unstructured data such as ratings, reviews, or images to profile the users for personalized recommendation. Though effective, it is difficult for existing approaches to model the explicit relationship between different information that we know about the users and items. 
%On the other hand, Collaborative filtering (CF) as a powerful recommender model has been widely used in many real-world systems.
%Traditional CF usually base itself on the user ratings. 
%However, with the ever prospering of Web 2.0, many applications have accumulated plenty of user behavior data, such as review text, click records, \emph{etc}, which is of great importance for more comprehensive user profiling and more accurate recommendations.
In this paper, we would like to ask a key question, i.e., \emph{``can we extend the power of CF upon large-scale structured user behaviors?''}. 
The main challenge to answer this question is how to effectively integrate different types of user behaviors and item properties, while preserving the internal relationship between them to enhance the final performance of personalized recommendation.

Fortunately, the emerging success on knowledge-base embeddings (KBE) may shed some light on this problem, where heterogeneous information can be projected into a unified low-dimensional embedding space. By encoding the rich information from multi-type user behaviors and item properties into the final user/item embeddings, we can enhance the recommendation performance while preserving the internal structure of the knowledge.

Equipping recommender systems with structured knowledge also helps the system to generate informed explanations for the recommended items. Researchers have shown that providing personalized explanations for the recommendations helps to improve the persuasiveness, efficiency, effectiveness, and transparency of recommender systems \cite{zhang2018explainable,zhang2014explicit,herlocker2000explaining, tintarev2007survey,bilgic2005explaining, cramer2008effects, tintarev2011designing}. By preserving the knowledge structure about users, items, and heterogenous entities, we can conduct fuzzy reasoning over the knowledge-base embeddings (KBE) to generate tailored explanations for each user.

%{\color{red}Some studies show observability and relevance are two important factors to attract users to recommended items \cite{ai2017learning, van2016learning}. Recommender systems and users perceive the relevance of an item differently and it is a big transparency gap between them. The consequence is even if one item which satisfies users search intent maybe is not perceived relevant for users. It means providing relevant items is not enough and recommender systems need to provide justification to assist users in making informed decision. \cite{herlocker2000explaining, tintarev2007survey} discuss that chance of user acceptance for recommended items will be improved by providing appropriate explanation. Providing explanation is along with some benefits like user satisfaction, improve user's experience and trust, system transparency and etc \cite{bilgic2005explaining, cramer2008effects, tintarev2011designing, zhang2014explicit} and consequently increases the performance and proficiency of recommender systems. One dilemma in recommendation systems is to have a solution that is both highly accurate and easily explainable \cite{zhang2014explicit}.}

%Maintaining the structures with knowledge base also helps to generate informed recommendation explanations by reasoning over the knowledge .

Inspired by the above motivations, in this paper, we design a novel explainable CF framework over knowledge graphs. The main building block is an integration of traditional CF with the learning of knowledge-base embeddings. More specifically, we first define the concept of user-item knowledge graph, which encodes our knowledge about the user behaviors and item properties as a relational graph structure. The user-item knowledge graph focuses on how to depict different types of user behaviors and item properties over heterogeneous entities in a unified framework. Then, we extend the design philosophy of CF to learn over the knowledge graph for personalized recommendation. For each recommended item, we further conduct fuzzy reasoning over the paths in the knowledge graph based on soft matching to construct personalized explanations.

\noindent
\textbf{Contributions.}%Is this a list or a paragraph? %Is the bold neccesary?
~The main contributions of this paper can be summarized as follows:

\vspace{6pt} $\bullet$ We propose to integrate heterogeneous multi-type user behaviors and knowledge of the items into a unified graph structure for recommendation.%which to the best of our knowledge, is the first time in the filed of recommendation community.

$\bullet$ Based on the user-item knowledge graph, we extend traditional CF to learn over the heterogeneous knowledge for recommendation, which helps to capture the user preferences more comprehensively.

$\bullet$ We further propose a soft matching algorithm to construct explanations regarding the recommended items by searching over the paths in the graph embedding space.

$\bullet$ Extensive experiments verify that our model can consistently outperform many state-of-the-art baselines on real-world e-commerce datasets.

\vspace{6pt} In the following part of the paper, we first present related work in Section \ref{sec:related}, and then provide the problem formalization in Section \ref{sec:formalization}. Section \ref{sec:model} goes over the model for CF over knowledge graphs, and in Section \ref{sec:explanation} the soft matching method for generating explanations is illustrated. Experimental setup and discussion of the results are provided in Section \ref{sec:setup} and Section \ref{sec:results}, respectively. We conclude the work and point out some of the future research directions in Section \ref{sec:conclusions}.
% At last the experiment, result and conclusion is presented.

%!TEX root=paper.tex

\section{Related Work}\label{sec:related}

Using knowledge base to enhance the performance of recommender system is an intuitive idea, which has attracted research attention since the very early stage of the recommender system community. For example, Burke \cite{burke1999integrating} and Trewin \cite{burke2000knowledge} discussed the strengths and weaknesses of both knowledge-based and CF-based recommender systems, and introduced the possibility of a hybrid recommender system that combines the two approaches. Ghani and Fano \cite{ghani2002building} presented a case study of a system that recommends items based on a custom-built knowledge base that consists of products and associated semantic attributes. 
Heitmann \cite{heitmann2012open} proposed to conduct cross-domain personalization and recommendation based on multi-source semantic interest graphs.

More recently, Musto et al. \cite{musto2016semantics} and Noia et al. \cite{noia2016sprank} conducted semantic graph-based recommendation leveraging linked open data as external knowledge, Oramas et al. \cite{oramas2017sound} adopted knowledge graphs to produce sound and music recommendations, and Catherine et al. \cite{catherine2017explainable} proposed to jointly rank items and knowledge graph entities using a personalized page-rank procedure to produce recommendations together with the explanations.

Though intuitive, the difficulty of reasoning over the hard-coded paths on heterogeneous knowledge graph prevents existing approaches from leveraging CF on very different entities and relations, which further makes it difficult to take advantage of the wisdom of crowd. 

Fortunately, recent advances on KBE shed light on this problem. In KBE, entities and relations are learned as vector representations, and the connectivity between entities under a certain relation can be calculated in a soft manner based on their representations. Early approaches to knowledge-base embedding are based on matrix factorization \cite{nickel2011three,singh2008relational} or non-parametric Bayesian % Is the bold neccesary? PLease confirm the whole paper.
 frameworks \cite{miller2009nonparametric,zhu2012max}. More recently, the advance of neural embedding methods led to a lot of neural KBE approaches \cite{bordes2011learning}. 
%For example, Bordes et al. \cite{bordes2011learning} proposed to embed symbolic representations from knowledge bases into a continuous vector space with a neural network architecture. 
Bordes et al. \cite{bordes2013translating} designed a translation-based embedding model (transE) to jointly model entities and relationships within a single latent space, which were later generalized into hyperplane translation (transH) \cite{wang2014knowledge} and translation in separate entity space and relation spaces (transR) \cite{lin2015learning}.

Researchers attempted to leverage knowledge base embeddings for recommendation. \linebreak{For example}, Zhang et al. \cite{zhang2016collaborativekdd} proposed collaborative KBE to learn the items' semantic representations from the knowledge base for recommendation, but the whole knowledge of an item is learned as a single item vector and the model did not preserve the knowledge-base structure for reasoning; He et al. \cite{he2017translation} leveraged translation-based embedding for recommendation by modeling items as entities and users as relations, but they did not consider the knowledge information of users and items for recommendation. 
Even though many studies have applied neural techniques for recommender systems~\cite{Li2015DeepCF, Wang:2015:CDL:2783258.2783273, Li:2017:CVA:3097983.3098077}, none of the previous work has leveraged KBE for explainable recommendation \cite{zhang2018explainable}. However, we will show that a great advantage of learning KBEs is to generate very straight forward explanations by soft-reasoning over the user-to-item paths.

%Researchers have shown that generating model-based explanations helps to improve the persuasiveness, efficiency, effectiveness, and transparency of recommender systems \cite{zhang2014explicit,herlocker2000explaining, tintarev2007survey,bilgic2005explaining, cramer2008effects, tintarev2011designing}.

%Researchers have shown that \cite{zhang2018explainable}

%Fortunately, recent years have witnessed the success of heterogenous knowledge base embedding techniques \cite{bordes2013translating,wang2014knowledge,lin2015learning}, which can help to learn the embeddings of very different entities to support various application scenarios such as question answering \cite{bordes2014question} and relation extraction from text \cite{lin2015learning}. To the best of our knowledge, this work is the first time to learn per-entity-level knowledge base embeddings for personalized recommendation.

%mostly focus on using knowledge bases in the background of content-based recommendation \cite{,szomszor2007folksonomies}, and they lack the ability of collaboratively leverage the wisdom of the crowd.

%!TEX root=paper.tex

\section{Problem Formulation}\label{sec:formalization}

In this paper, we focus on \textit{explainable product recommendation}, where the objective of the recommender system is to recommend products to users and explain why the products are recommended.
%We refer to this as the problem of .
%Thus, for explainable recommendation, we need to jointly model users, items and related product knowledge in a single framework.   
% product recommendation -> recommend purchase item
% based on rating
% not enough for explanation -> knowledge graph
%Traditional recommendation problems mainly concerns about the relationship between users and products.

%Explainable product recommendation, however, requires the recommender system to understand the user-item relationship as well as other related knowledge. % in order to generate recommendation explanations that are both readable and reasonable. 
Formally, we first construct a knowledge-base as a set of triplets $S = \{(e_h, e_t, r)\}$ for recommendation, where $e_h$ is a head entity, $e_t$ is a tail entity, and $r$ is the relationship from $e_h$ to $e_t$.
Then the goal of explainable recommendation is twofold, 1) for each user $u$, find one or a set of items $i$ that are most likely to be purchased by the user, and 2) for each retrieved user-item pair, construct a natural language sentence based on $S$ to explain why the user should purchase the item.
%As discussed previously, e-shopping websites often contain heterogeneous information about each product including both unstructured (e.g. review text) and structured (e.g. brands) data.

\noindent
For simplicity, we consider 5 types of entities (i.e., $e_h$ or $e_t$) for explainable recommendation:

\vspace{6 pt}
%\begin{itemize}
 $\bullet${ \textit{user}}%Is the italic neccessary? Please confirm the whole paper.
: the users of the recommender system.

$\bullet$ \textit{item}: the products in the system to be recommended.

$\bullet$ \textit{word}: the words in product names, descriptions or reviews.

$\bullet$ \textit{brand}: the brand/producers of the product.

$\bullet$ \textit{category}: the categories that a product belongs to. 
%\end{itemize}
\vspace{6 pt}

\noindent
\vspace{6 pt} Also, we consider 6 types of relationships (i.e., $r$) between entities:

$\bullet$ \textit{Purchase}: the relation from a user to an item, which means that the user has bought the item.

$\bullet$ \textit{Mention}: the relation from a user or an item to a word, which means the word is mentioned in the user's or item's reviews.

$\bullet$ \textit{Belongs\_to}%Is this should be a hyphen? Please confirm the whole paper. 
: the relation from an item to a category, which means that the item belongs to \linebreak{the category.}

$\bullet$ \textit{Produced\_by}: the relation from an item to a brand, which means that the item is produced by \linebreak{the brand.}

$\bullet$ \textit{Bought\_together}: the relation from an item to another item, which means that the items have been purchased together in a single transaction.

$\bullet$ \textit{Also\_bought}: the relation from an item to another item, which means the items have been purchased by same users.

$\bullet$ \textit{Also\_viewed}: the relation from an item to another item, which means that the second item was viewed before or after the purchase of the first item in a single session.

\vspace{6 pt} Therefore, for explainable product recommendation, the first goal is to retrieve item $i$ that are likely to have the \textit{Purchase}
 relationship with user $u$, and the second goal is to provide explanations for the $(u,i)$ pair based on the relations and entities related to them.
%provide explanations for a set of items $i$ that are likely to have the relationship %\textit{Purchase} with the current user $u$.

% in product data, there are multiple data and relationships
% Here we consider xxx

%The effectiveness of product recommendation and explanation generation with knowledge graph is mostly unexplored.
% Two research questions
% How to recommend
% How to explain

%\begin{figure*}[t]
%%\begin{figure}[H]
%%	\centering
%%	\vspace{-5pt}
%%	\includegraphics[width=5in]{fig/graph_illustration.pdf}
%%	\caption{The construction process of knowledge graph with our model. Each entity is represented with a latent vector, and each relation is modeled as a linear translation from one entity to another entity parameterized by the relation embedding.}
%%	\vspace{-5pt}
%%	\label{fig:illustration}
%%%\end{figure*}
%%\end{figure}

\section{Collaborative Filtering on Knowledge Graphs}\label{sec:model}

We now describe our model for explainable recommendation.
Our model is a CF model built on user-item knowledge graph.
In this section, we first introduce how to model the entities and relations as a product knowledge graph, and then we discuss how to optimize the model parameters for recommendation. 

%the details of the proposed collaborative filtering method on knowledge graph for explainable product recommendation
%Despite the extensive studies on product recommendation~\cite{xxx}, as far as we know, there is no existing work that has successfully unified the modeling of unstructured text data with structured product knowledge graph for explainable recommendation.
%In this section, we propose an embedding model to conduct collaborative filtering on knowledge graph for recommendation.
%Therefore, the questions of how to recommend products using product knowledge and how to explain %recommendation results using knowledge graph are mostly unexplored.

%In this section, we describe the problem of collaborative filtering on knowledge graph and propose a 

\subsection{Relation Modeling as Entity Translations}\label{sec:trans}

As discussed previously, we assume that the product knowledge can be represented as a set of triplets $S = \{(e_h, e_t, r)\}$, where $r$ is the relation from entity $e_h$ to entity $e_t$.
Because an entity can be associated with one or more other entities through a single or multiple relations, we propose to separate the modeling of entity and relation for CF.
Specifically, we project each entity to a low-dimensional latent space and treat each relation as a translation function that converts one entity to another.
Inspired by ~\cite{bordes2013translating}, we represent $e_h$ and $e_t$ as latent vectors $\bm{e_h} \in \mathbb{R}^{d}$ and $\bm{e_t} \in \mathbb{R}^{d}$, and model their relationship $r$ as a linear projection from $e_h$ to $e_t$ parameterized by $\bm{r} \in \mathbb{R}^{d}$, namely,
\begin{equation}
\bm{e_t} = trans(e_h, r) = \bm{e_h} + \bm{r} 
\label{equ:trans}
\end{equation}
%where $d$ is the number of dimension (embedding size) for the latent vectors. 

\begin{figure}[H]
	\centering
	\vspace{-5pt}
	\includegraphics[width=5in]{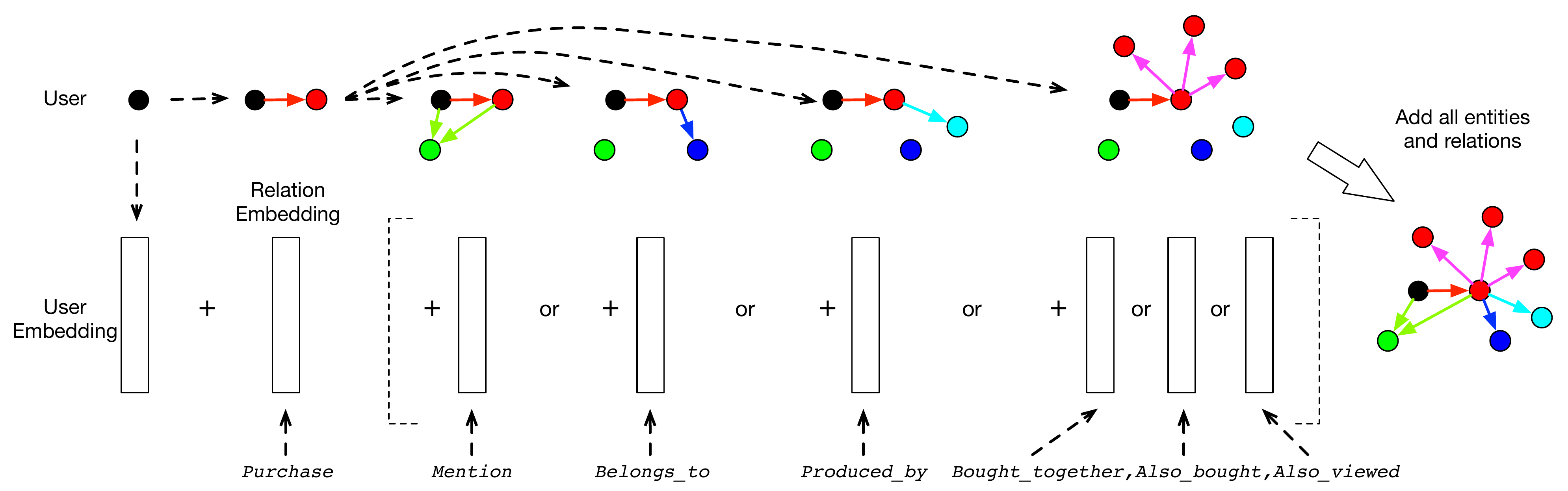}
	\caption{The construction process of knowledge graph with our model. Each entity is represented with a latent vector, and each relation is modeled as a linear translation from one entity to another entity parameterized by the relation embedding.}%Please check and confirm.
	\vspace{-5pt}
	\label{fig:illustration}
%\end{figure*}
\end{figure}

To learn the entity embeddings, %Please check and confirm.
we can construct a product knowledge graph by linking entities with the translation function in the latent space.
An example generation process of such a graph is shown in Figure~\ref{fig:illustration}.
%As shown in Figure~\ref{fig:illustration}, we can construct a product knowledge graph based on the embedding representations of entities and relations. 

Solving Equation~(\ref{equ:trans}) for all $(e_h, e_t, r) \in S$, however, is infeasible in practice.
On one hand, a trivial solution that constructs a single latent vector for all entities with the same type will lead to inferior recommendation performance as it ignores the differences between users and items.
On the other hand, deriving a solution that assigns different latent vectors for all entities in $S$ is mathematically impossible because an entity can be linked to multiple entities with a single relationship.
For example, we cannot find a single vector for \textit{Also\_viewed} that translates an item to multiple items that have different \linebreak{latent representations. }

To solve the problems, we propose to relax the constrains of Equation~(\ref{equ:trans}) and adopt an embedding-based generative framework to learn it.
Empirically, we want the translation model $trans(e_h, r) \approx \bm{e_t}$ for an observed relation triplet $(e_h, e_t, r) \in S$ and $trans(e_h, r) \neq \bm{e_t'}$ for an unobserved triplet $(e_h, e_t', r) \notin S$.
In other words, we want $trans(e_h, r)$ to assign high probability for observing $e_t$ but low probability for observing $e_t'$, which is exactly the goal of the embedding-based generative framework. 
%Therefore, we adopt an embedding-based generative framework to learn the distributed representation of entities and relationships.
The embedding-based generative framework is first proposed by Mikolov et al.~\cite{mikolov2013distributed} and has been widely used in word embedding \cite{mikolov2013distributed,le2014distributed}, recommendation \cite{zhang2017joint,zheng2017joint}, and information retrieval tasks~\cite{ai2016analysis, ai2017learning}.
Formally, for an observed relation triplet $(e_h, e_t, r) \in S$, we can learn the translation model $trans(e_h, r)$ by optimizing the generative probability of $e_t$ given $trans(e_h, r)$, which is defined as:
\begin{equation}
P(e_t | trans(e_h, r)) = \frac{\exp(\bm{e_t} \cdot trans(e_h, r))}{\sum_{e_t' \in E_t}\exp(\bm{e_t'} \cdot trans(e_h, r))}
\label{equ:softmax}
\end{equation}
where $E_t$ is the set of all possible entities that share the same type with $e_t$.

Because %Please check and confirm.
 Equation~(\ref{equ:softmax}) is a softmax function of $e_t$ over $E_t$, the maximization of $P(e_t | trans(e_h, r))$ will explicitly increase the similarity of $trans(e_h, r)$ and $e_t$ but decease the similarity between $trans(e_h, r)$ and other entities. 
In this way, we convert Equation~(\ref{equ:trans}) into an optimization problem that can be solved with iterative optimization algorithms such as gradient decent.
Another advantage of the proposed model is that it provides a theoretically principled method to conduct soft match between tail entities and the translation model. 
This is important for the extraction of recommendation explanations, which will be discussed in Section~\ref{sec:explanation}.

\subsection{Optimization Algorithm}

For model optimization, we learn the representations of entities and relations by maximizing the likelihood of all observed relation triplets. 
Let $S$ be the set of observed triplets $(e_h, e_t, r)$ in the training data, then we can compute the likelihood of $S$ defined as
\begin{equation}
\mathcal{L}(S) = \log\!\!\!\!\prod_{(e_h, e_t, r) \in S}\!\!\!\! P(e_t | trans(e_h, r))
\label{equ:log_likelihood}
\end{equation}
where $P(e_t | trans(e_h, r))$ is the posterior probability of $e_t$ computed with Equation~(\ref{equ:softmax}).

The computation cost of $\mathcal{L}(S)$, however, is prohibitive in practice because of the softmax function.
For efficient training, we adopt a negative sampling strategy to approximate $P(e_t | trans(e_h, r))$~\cite{mikolov2013efficient}.
Specifically, for each observed relation triplet $(e_h, e_t, r)$, we randomly sample a set of ``negative'' entities with the same type of $e_t$. % to approximate the denominator of the softmax function. 
Then the log likelihood of $(e_h, e_t, r)$ is approximated as
\begin{equation}
\begin{split}
\log P(e_t | trans(e_h, r)) \approx \log \sigma(\bm{e_t} \cdot trans(e_h, r)) + k\cdot \mathbb{E}_{e_t'\sim P_t}[\log\sigma(-\bm{e_t'} \cdot trans(e_h, r))]
\end{split}
\label{equ:negative_sample}
\end{equation}
where $k$ is the number of negative samples, $P_t$ is a predefined noisy distribution over entities with the type of $e_t$, and $\sigma(x)$ is a sigmoid function as $\sigma(x) = \frac{1}{1 + e^{-x}}$.
Therefore, $\mathcal{L}(S)$ can be reformulated as the sum of the log-likelihood of $(e_h, e_t, r)\in S$ as
\begin{equation}
\begin{split}
\mathcal{L}(S) = \!\!\!\!\sum_{(e_h, e_t, r) \in S}\!\!\!\! \log \sigma(\bm{e_t} \cdot trans(e_h, r)) +k\cdot \mathbb{E}_{e_t'\sim P_t}[\log\sigma(-\bm{e_t'} \cdot trans(e_h, r))]
\end{split}
\label{equ:final_loss}
\end{equation}
%where the embeddings of all entities $\{\bm{e}\}$ and relationships $\{\bm{r}\}$ appeared in $S$ are the parameters to learn in the training process.
%The final optimization algorithm is described in Algorithm~\ref{alg:A}.

We also tested the $\ell_2$-norm loss function used in TransE model and it does not provide any improvement compared to our inner product-based model with log-likelihood loss, and it is also difficulty for $\ell_2$-norm loss to generate expansions, as a result, we adopt our loss function for embedding, recommendation, and explanation in this work. To better illustrate the relationship between our model and a traditional CF method based on matrix factorization, we conduct the following analysis. 
Inspired by ~\cite{levy2014neural}, we derive the local objective for the maximization of Equation~(\ref{equ:final_loss}) on a specific relation triplet $(e_h, e_t, r)$:
\begin{equation}
\begin{split}
\ell(e_h, e_t, r) = \#(e_h, e_t, r)\cdot\log\sigma(\bm{e_t} \cdot trans(e_h, r)) + k \cdot \#(e_h,r) \cdot P_t(e_t)\cdot \log\sigma(-\bm{e_t} \cdot trans(e_h, r))
\end{split}
\label{equ:local_object}
\end{equation} 
where $\#(e_h, e_t, r)$ and $\#(e_h,r)$ are the frequency of $(e_h, e_t, r)$ and $(e_h,r)$ in the training data.
If we further compute the partial derivative of $\ell(e_h, e_t, r)$ with respect to $x = \bm{e_t} \cdot trans(e_h, r)$, we have
\begin{equation}
\frac{\partial \ell(e_h, e_t, r)}{\partial x} = \#(e_h, e_t, r)\cdot \sigma(-x) - k\cdot \#(e_h,r)\cdot P_t(e_t)\cdot \sigma(x)
\label{equ:derivative}
\end{equation}

When the training process has converged, the partial derivative of $\ell(e_h, e_t, r)$ should be 0, and then we~have
\begin{equation}
\begin{split}
x = \bm{e_t} \cdot trans(e_h, r) & = \log(\frac{\#(e_h, e_t, r)}{\#(e_h,r)}\cdot \frac{1}{P_t(e_t)}) - \log k %\\
\end{split}
\label{equ:l_final_obj}
\end{equation}

As we can see, the left-hand side  is the product of the latent vectors for $e_t$ and $trans(e_h,r)$; and the right-hand side of Equation~(\ref{equ:l_final_obj}) is a shifted version of the pointwise mutual information between $e_t$ and $(e_h, r)$.
Therefore, maximizing the log likelihood of observed triplet set $S$ with negative sampling is actually factorizing the matrix of mutual information between the head-tail entity pairs $(e_h, e_t)$ of relation $r$.
From this perspective, our model is a variation of factorization methods that can jointly factorize multiple relation matrix on a product knowledge graph.
%has much more flexibility and extendability in terms of model design and parameter optimizations.

As shown in Equation~(\ref{equ:l_final_obj}), the final objective of the proposed model is controlled by the noisy distribution $P_t$.
%It penalizes the mutual information between $(e_h, r)$ with $e_t$ that has high noise $P_t(e_t)$.
Similar to previous studies~\cite{mikolov2013efficient,le2014distributed,ai2016analysis}, we notice that the relationships with tail entities that have high frequency in the collection reveal less information about the properties of the head entity.
Therefore, we define the noisy probability $P_t(e_t)$ for each relation $r$ (except \textit{Purchase}) as the frequency distribution of $(e_h, e_t, r)$ in $S$ so that the mutual information on frequent tail entities will be penalized in optimization process.
For \textit{Purchase}, however, we define $P_t$ as a uniform distribution to avoid unnecessary biases toward certain items.\vspace{12 pt}

\begin{algorithm}[H]
	\small
	\caption{Recommendation Explanation Extraction}
	\SetAlgoLined
	\label{alg:REE}
	\SetKwFunction{algo}{Main} \SetKwFunction{proc}{BFS}
	\KwIn{$S=\{(E_h, E_t, r)\}$, $e_u$, $e_i$, maximum depth $z$}
	\KwOut{$e_x, R_{\alpha}, R_{\beta}$}
	\SetKwProg{myalg}{Procedure}{}{}
	\myalg{\algo{}}{
		\nl $V_u, P_u, \mathcal{R}_u = BFS(S, e_u, z)$. \\
		\nl $V_i, P_i, \mathcal{R}_i = BFS(S, e_i, z)$. \\
		\nl $P \leftarrow \{\}$. \\
		\nl \For{$e \in V_u \cap V_i$}{
			\nl $P[e] = P_u(e) \cdot P_i(e)$.
		}
		\nl Pick up $e_x \in V_u \cap V_i$ with the largest $P[e]$. \\
		\nl $R_{\alpha} = \mathcal{R}_u[e_x], R_{\beta} = \mathcal{R}_i[e_x]$. \\
		\nl \Return $e_x, R_{\alpha}, R_{\beta}$
	}
	\SetKwProg{myproc}{Function}{}{}
	\myproc{\proc{$S, e, z$}}{
		\nl $V_e \leftarrow $ all entities in the entity set $E_t$ within $z$ hops from $e$. \\
		\nl $P_e \leftarrow $ the probability of each entity in $V_e$ computed by Eq~(\ref{equ:one_direction_prob}).\\
		\nl $\mathcal{R}_e \leftarrow $ the paths from $e$ to the space of each entity in $V_e$.\\
		\nl \KwRet $V_e, P_e, \mathcal{R}_e$;}
\end{algorithm}

\section{Recommendation Explanation with Knowledge Reasoning}\label{sec:explanation}

In this section, we describe how to create recommendation explanations with the proposed model.
We first introduce the concept of explanation path and describe how to generate natural language explanations with it.
Then we propose a soft matching algorithm to find explanation path for any user-item pair in the latent space.

An overview of our algorithm is shown in Algorithm~\ref{alg:REE}.
In the algorithm, we first conduct breath first search (BFS) with maximum depth $z$ from the user $e_u$ and the item $e_i$ to find an explanation path that can potentially link them. 
We memorize the paths and compute the path probability with soft matching (Equation~(\ref{equ:one_direction_prob}) and (\ref{equ:path_prob})).
%After that, we match the potential link entities from $e_u$ ($V_u$) and $e_i$ ($V_i$), and compute the probability of the corresponding explanation path with Equation~(\ref{equ:path_prob}).
Finally, we rank the explanation paths by their probabilities and return the best path to create the natural language explanation. In this work, we take $z$ as a hyper-parameter and tune the parameter by increasing its value in the experiments to search for non-empty intersections.

\subsection{Explanation Path}~\label{sec:path}

The key to generate an explanation of the recommendation is to find a sound logic inference sequence from the user to the item in the knowledge graph.
In this work, we propose to find such a sequence by constructing an explanation path between the user and the item in the latent knowledge~space.

Formally, let $E_h^r$ and $E_t^r$ be the sets of all possible head entities and tail entities for a relation $r$. 
We define an explanation path from entity $e_u$ to entity $e_i$ as two sets of relation $R_{\alpha} = \{r_{\alpha} | \alpha \in [1,m]\}$ and $R_{\beta} = \{r_{\beta} | \beta \in [1,n]\}$ such that
\begin{equation}
\begin{split}
\bm{e_u} + \sum_{\alpha=1}^m \bm{r_{\alpha}} = \bm{e_i} + \sum_{\beta=1}^n \bm{r_{\beta}}
\end{split}
\label{equ:path}
\end{equation}
where $e_u \!\in\! E_h^{r_{\alpha}}$ for $\alpha\!=\!1$, $e_i \!\in\! E_h^{r_{\beta}}$ for $\beta\!=\!1$, $E_t^{r_{\alpha-1}} \!=\! E_h^{r_{\alpha}}$ for $\alpha \!\in\! [2,m]$, $E_t^{r_{\beta-1}}\! =\! E_h^{r_{\beta}}$ \linebreak{for $\beta\! \in\! [2,n]$, and} $E_t^{r_{\alpha}} \!=\! E_t^{r_{\beta}}$ for $\alpha\!=\!m, \beta\!=\!n$.
In other words, there is an explanation path between $e_u$ and $e_i$ if there is an entity that can be inferred by both $e_u$ (with $R_{\alpha}$) and $e_i$ (with $R_{\beta}$) with the observed relations in the knowledge graph.

%%\begin{figure}[t!]
%%\begin{figure}[H]
%%	\centering
%%	\includegraphics[width=0.5\linewidth, ]{fig/explanation_path.pdf}
%%	\caption{Example explanation paths between a user \textit{Bob} and a recommended item \textit{iPad} in the product knowledge graph.}		\vspace{-10pt}
%%	\label{fig:example_path}
%%\end{figure}

\begin{figure}[H]
	\centering
	\includegraphics[width=0.5\linewidth, ]{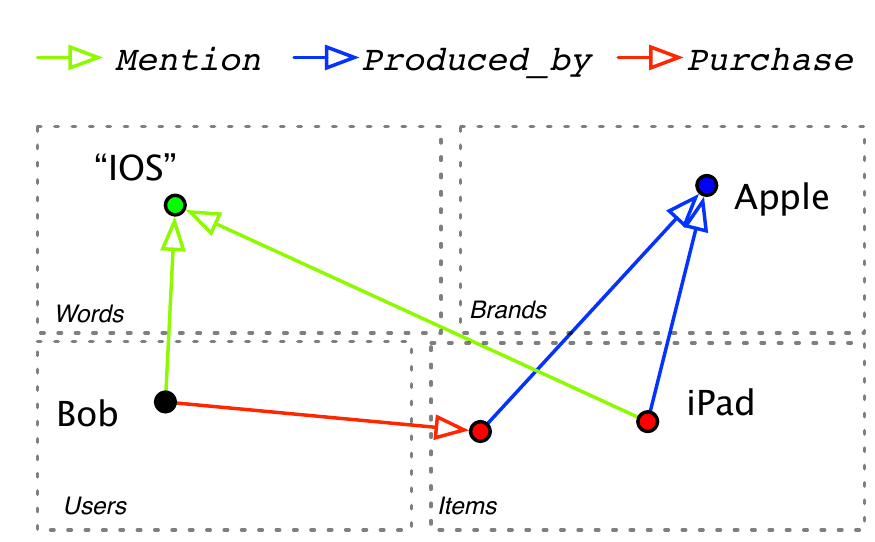}
	\caption{Example explanation paths between a user \textit{Bob} and a recommended item \textit{iPad} in the product knowledge graph.}		\vspace{-10pt}
	\label{fig:example_path}
\end{figure}

For better illustration, we depict a recommendation example where an item (\textit{iPad}) is recommended to a user (\textit{Bob}) in Figure~\ref{fig:example_path}.
As we can see, the word ``IOS'' can be inferred by \textit{iPad} and \textit{Bob} using the relation \textit{Mention}; the brand \textit{Apple} can be inferred by \textit{iPad} using \textit{Produced\_by}, and by \textit{Bob} using \textit{Purchase} + \textit{Produced\_by}.
Thus, we have two explanation paths that link \textit{Bob} with \textit{iPad}.
%To actually explain why \textit{iPad} should be recommended to \textit{Bob}, we can adopt simple templates on the explanation paths to generate natural language explanations.

To generate explanations, we can create simple templates base on the relation type and apply them to the explanation paths.
For example, we can say that \textit{Bob} may be interested in \textit{iPad} because he often \textit{mentions} ``IOS'' in his reviews, and ``IOS''%Please define, if appropriate.
is often \textit{mentioned} in the reviews of \textit{iPad}; or that \textit{Bob} may be interested in \textit{iPad} because he often \textit{purchases} products \textit{produced by} \textit{Apple}, and \textit{iPad} is also \textit{produced} by \textit{Apple}. In these explanations, the italic words are entities and relations on the explanation path.

\subsection{Entity Soft Matching}

Finding a valid explanation path with observed relations, however, is often difficult for an arbitrary user-item pair.
In practice, product knowledge graphs tend to be sparse.
For example, \linebreak{the density} of user-item matrix in Amazon review datasets is usually below 0.1\%~\cite{mcauley2015image}.
\linebreak{Thus, the} available relation triplets in the observed data are limited.
To solve the problem, we propose to conduct entity soft matching in the latent space for explanation construction.

As discussed in Section~\ref{sec:trans}, we learn the distributed representations of entities and relations by optimizing the generative probability of observed relation triplets in Equation~(\ref{equ:softmax}). 
Thus, we can extend the softmax function to compute the probability of entity $e_x \in E_t^{r_{m}}$ given $e_u$ and the relation set $R_{\alpha} = \{r_{\alpha} | \alpha \in [1,m]\}$ as
\begin{equation}
\begin{split}
P(e_x | trans(e_u, R_{\alpha})) = \frac{\exp(\bm{e_x} \cdot trans(e_u, R_{\alpha}))}{\sum_{e' \in E_t^{r_{m}}}\exp(\bm{e'} \cdot trans(e_u, R_{\alpha}))}
\end{split}
\label{equ:one_direction_prob}
\end{equation} 
where $E_t^{r_{m}}$ is the tail entity set of $r_{m}$, and $trans(e_u, R_{\alpha}) = \bm{e_u} + \sum_{\alpha=1}^m \bm{r_{\alpha}}$.
Therefore, we can construct an explanation path for an arbitrary user $e_u$ and item $e_i$ with relation sets $R_{\alpha} = \{r_{\alpha} | \alpha \in [1,m]\}$ and $R_{\beta} = \{r_{\beta} | \beta \in [1,n]\}$ through any intermediate $e_x \in E_t^{r_{m}}$, and compute the probability of this explanation path as:
%We can any $e_x$ to link the 
%As a result, we can soft match any $e_x$ with an arbitrary user $e_u$ and item $e_i$ through the relation sets $R_{\alpha} = \{r_{\alpha} | \alpha \in [1,m]\}$ and $R_{\beta} = \{r_{\beta} | \beta \in [1,n]\}$, and construct an explanation path with the probability of
\begin{equation}
P(e_x|e_u,R_{\alpha}, e_i,R_{\beta}) \!=\! P(e_x | trans(e_u, \!R_{\alpha}))  P(e_x | trans(e_i, \!R_{\beta}))
\label{equ:path_prob}
\end{equation}

To find the best explanation for $(e_u, e_i)$, we can rank all paths by $P(e_x|e_u,R_{\alpha}, e_i,R_{\beta})$ and pick up the best one to generate natural language explanations with predefined templates. It should be noted that learning with single hops may not guarantee the quality of multiple hops matching, but it also significantly simplifies the design of the training algorithm and increases the generalizability of the model, and helps to generate explanations more easily. Besides, using the single-hop training strategy has already been also to compete with many of the baselines. However, we believe that model training with multiple hops directly is a promising problem and we will design new models for this problem as a future work.

\section{Experimental Setup}\label{sec:setup}

In this section, we introduce the test bed of our experiments and discuss our evaluation settings in details.

\subsection{Datasets}
%\footnote{\url{http://jmcauley.ucsd.edu/data/amazon/}}
We conducted experiments on the Amazon review dataset~\cite{mcauley2015image}, which contains product reviews in 24 categories on Amazon.com and rich metadata such as prices, brands, etc.
Specifically, we used the 5-core data of \textit{CDs and Vinyl}, \textit{Clothing}, \textit{Cell Phones}, and \textit{Beauty}, in which each user or item has at least 5 associated reviews.
%Because a user can only write a review for an item after purchasing it, we extract purchased user-item pairs directly based on the review information.

Statistics about entities and relations used in our experiments are shown in Table~\ref{tab:dataset_statistics}. 
\linebreak{Overall, the} interactions between users, items and other entities are highly sparse.
% (e.g. the density of user-item purchase matrix is lower than 0.1\% for all datasets). 
For each dataset, we randomly sampled 70\% of user purchase as the training data and used the rest 30\% as the test set. 
This means that each user has at least 3 reviews observed in the training process and 2 reviews hidden for evaluation purposes.
Thus, the objective of product recommendation is to find and recommend items that are purchased by the user in the test set.

\begin{table}[H]
%%\begin{table*}[t]
	\centering
	\small
	\setlength{\tabcolsep}{2pt}
	\caption{Statistics of the 5-core datasets for \textit{CDs \& Vinyl}, \textit{Clothing}, \textit{Cell Phones \& Accessories}, and \textit{Beauty} in Amazon. 
	}
	\vspace{-5pt}
	\begin{tabular}{ l c  c  c  c   } %p{5mm}
		\toprule
		& \textit{\bf{CDs \& Vinyl}} & \textit{\bf{Clothing}} & \textit{\bf{Cell Phones \& Accessories}} & \textit{\bf{Beauty}}\\
		\midrule
		\textbf{Entities}\\
		~~~~\#Reviews & 1,097,591 & 278,677 & 194,439 & 198,502\\ %
		~~~~\#Words per review & 174.57 $\pm$ 177.05 & 62.21 $\pm$ 60.16 & 93.50 $\pm  $131.65 & 90.90 $\pm$ 91.86\\
		~~~~\#Users & 75,258 & 39,387 & 27,879 & 22,363 \\ %
		~~~~\#Items & 64,443 & 23,033 & 10,429 & 12,101\\ %
		~~~~\#Brands & 1414 & 1182 & 955 & 2077\\ %
		~~~~\#Categories & 770 & 1,193 & 206 & 248\\ %
%		~~~~\#Sparsity & 99.977\% & 99.969\% & 99.933\% & 99.926\%\\%
		~~~~Density & 0.023\% & 0.031\% & 0.067\% & 0.074\%\\%
		\midrule
		\textbf{Relations}\\
		~~~~\#\textit{Purchase} per user & 14.58 $\pm$ 39.13 & 7.08 $\pm$ 3.59 & 6.97 $\pm$ 4.55 & 8.88 $\pm$ 8.16\\ %
		~~~~\#\textit{Mention} per user & 2545.92 $\pm$ 10,942.31 & 440.20 $\pm$ 452.38 & 652.08 $\pm$ 1335.76 & 806.89 $\pm$ 1344.08\\ %
		~~~~\#\textit{Mention} per item & 2973.19 $\pm$ 5490.93 & 752.75 $\pm$ 909.42 & 1743.16 $\pm$ 3482.76 & 1491.16 $\pm$ 2553.93\\ %
		~~~~\#\textit{Also\_bought} per item & 57.28 $\pm$ 39.22 & 61.35 $\pm$ 32.99 & 56.53 $\pm$ 35.82 & 73.65 $\pm$ 30.69\\ %
		~~~~\#\textit{Also\_viewed} per item & 0.27 $\pm$ 1.86 & 6.29 $\pm$ 6.17 & 1.24 $\pm$ 4.29 & 12.84 $\pm$ 8.97\\ %
		~~~~\#\textit{Bought\_together} per item & 0.68 $\pm$ 0.80 & 0.69 $\pm$ 0.90 & 0.81 $\pm$ 0.77 & 0.75 $\pm$ 0.72\\ %
		~~~~\#\textit{Produced\_by} per item & 0.21 $\pm$ 0.41 & 0.17 $\pm$ 0.38 & 0.52 $\pm$ 0.50 & 0.83 $\pm$ 0.38\\ %
		~~~~\#\textit{Belongs\_to} per item & 7.25 $\pm$ 3.13 & 6.72 $\pm$ 2.15 & 3.49 $\pm$ 1.08 & 4.11 $\pm$ 0.70\\ %		

		%\midrule
		%\textbf{Train/Test}\\
		%~~~~Number of reviews & 1,275,432/413,756 & 720,006/262,612 & 804,090/293,501 & 150,048/44,391\\ %
		%~~~~Number of queries & 904/85 & 3313/1290 & 534/160 & 134/31\\
		%~~~~Number of user-query pairs & 1,204,928/5,505 & 1,490,349/232,668 & 1,287,214/45,490 & 114,177/665\\ %qrel
		%~~~~Relevant items per pair & 1.12$\pm$0.48/1.01$\pm$0.09 & 1.87$\pm$3.30/1.48$\pm$1.94 & 2.57$\pm$6.59/1.30$\pm$1.19 & 1.52$\pm$1.13/1.00$\pm$0.05\\ %qrel
		\bottomrule
	\end{tabular}
	\vspace{-5pt}
	\label{tab:dataset_statistics}
%%\end{table*}
\end{table}

\subsection{Evaluation}

To verify the effectiveness of the proposed model, we adopt six representative and state-of-the-art methods as baselines for performance comparison. 
Three of them are traditional recommendation methods based on matrix factorization (BPR~\cite{bpr}, BPR-HFT~\cite{mcauley2013hidden}, and VBPR~\cite{he2016vbpr}), and the other three are deep models for product recommendation (DeepCoNN~\cite{zheng2017joint}, CKE~\cite{zhang2016collaborativekdd}, and JRL~\cite{zhang2017joint}). 

\vspace{6 pt} $\bullet~$\textbf{BPR: } The Bayesian personalized ranking~\cite{bpr} model is a popular method for top-N recommendation that learns latent representations of users and items by optimizing the pairwise preferences between different user-item pairs. 
In this paper, we adopt matrix factorization as the prediction component for BPR.

$\bullet~$\textbf{BPR-HFT: } The hidden factors and topics (HFT) model~\cite{mcauley2013hidden} integrates latent factors with topic models for recommendation.
The original HFT model is optimized for rating prediction tasks.
For fair comparison, we learn the model parameters under the pairwise ranking framework of BPR for \linebreak{top-N recommendation.}

$\bullet~$\textbf{VBPR: } The visual Bayesian personalized ranking~\cite{he2016vbpr} model is a state-of-the-art method that incorporate product image features into the framework of BPR for recommendation.

$\bullet~$\textbf{TransRec: } The translation-based recommendation approach proposed in ~\cite{he2017translation}, which takes items as entities and users as relations, and leveraged translation-based embeddings to learn the similarity between user and items for personalized recommendation. We adopted $L_2$ loss function, which was reported to have better performance in \cite{he2017translation}. 
Notice that TransRec is different from our model because our model treats both items and users as entities, and learns embedding representations for different types of knowledge (e.g., brands, categories) as well as their relationships. 

$\bullet~$\textbf{DeepCoNN: } The Deep Cooperative Neural Networks model for recommendation~\cite{zheng2017joint} is a neural model that applies a convolutional neural network (CNN) over the textual reviews to jointly model users and items for recommendation.

$\bullet~$\textbf{CKE: } The collaborative KBE model is a state-of-the-art neural model~\cite{zhang2016collaborativekdd} that integrates text, images, and knowledge base for recommendation. It is similar to our model as they both use text and structured product knowledge, but it builds separate models on each type of data to construct item representations while our model constructs a knowledge graph that jointly embeds all entities and~relations. 

$\bullet~$\textbf{JRL: } The joint representation learning model~\cite{zhang2017joint} is a state-of-the-art neural recommender, which leverage multi-model information including text, images and ratings for \linebreak{Top-N recommendation.}%Is the capital neccesary? PLease confirm the whole paper.

\vspace{6 pt} The performance evaluation is conducted on the test set where only purchased items are considered to be relevant to the corresponding user.
Specifically, we adopt four ranking measures for top-N recommendation, which are the Normalized Discounted Cumulative Gain (\textbf{NDCG}), Precision (\textbf{Prec.}), \textbf{Recall}, and the percentage of users that have at least one correct recommendation\linebreak{ (Hit-Ratio,} \textbf{HR}).
All ranking metrics are computed based on the top-10 results for each test user.
Significant test is conducted based on the Fisher randomization test~\cite{smucker2007comparison}.

\subsection{Parameter Settings}

Our model is trained with stochastic gradient decent on a Nvidia Titan X GPU.
We set the initial learning rate as 0.5 and gradually decrease it to 0.0 during the training process. 
We set the batch size as 64 and clip the norm of batch gradients with 5.
For each dataset, we train the model for 20 epochs and set the negative sampling number as 5.
We tune the dimension of embeddings from 10 to 500\linebreak{ ([10, 50, 100, 200, 300, 400, 500]) }and report the best performance of each model in Section~\ref{sec:results}.

We also conduct five-fold cross-validation on the training data to tune the hyper-parameters for baselines. 
For BPR-HFT, the best performance is achieved when the number of topics is 10.
For BPR and VBPR, the regularization coefficient $\lambda =10$ worked the best in most cases.
Similar to our model, we tune the number of latent factors (the embedding size) from 10 to 500 and only report the best performance of each baseline.

\section{Results and Discussion}\label{sec:results}

In this section, we discuss the experimental results. 
We first compare the performance of our model with the baseline methods on top-N recommendation.
Then we conduct case studies to show the effectiveness of our model for recommendation explanation. 

\subsection{Recommendation Performance}

\noindent
Our experiments mainly focus on two research questions:

\vspace{6 pt} 

\hspace{-1cm}$\bullet$ \textbf{RQ1}: Does incorporating knowledge-base in our model produce better recommendation performance?

\hspace{-1cm}$\bullet$ \textbf{RQ2}: Which types of product knowledge are most useful for top-N recommendation? 

\hspace{-1cm}$\bullet$ \textbf{RQ3}: What is the efficiency of our knowledge-enhanced recommendation model compared to other algorithms?

% research questions
% effectiveness of product knowledge
% what is the important information

\vspace{6 pt} To answer \textbf{RQ1}, we report the results of our model and the baseline methods in Table~\ref{tab:result}.
\linebreak{As shown} in Table~\ref{tab:result}, the deep models with rich auxiliary information (DeepCoNN, CKE, and JRL) perform better in general than the shallow methods (BPR, BPR-HFT, VBPR, TransRec) on most datasets, which is coherent with previous studies~\cite{zhang2016collaborativekdd,zheng2017joint,zhang2017joint}.
%This is coherent with previous~\cite{zhang2016collaborativekdd,zheng2017joint,zhang2017joint}. 
%This demonstrates the effectiveness of neural recommender systems.
Among different neural baselines, JRL obtains the best performance in our experiments.
It produced 80\% or more improvements over the matrix factorization baselines and 10\% or more over the other deep recommendation models.
Overall, our model outperformed all the baseline models consistently and significantly. 
It obtained 5.6\% NDCG improvement over the best baseline (i.e., JRL) on \textit{CDs and Vinyl}, 78.16\% on \textit{Clothing}, 23.05\% on \textit{Cell Phones}, and 45.56\% on \textit{Beauty}.
This shows that the proposed model can effectively incorporate product knowledge graph and is highly competitive for top-N recommendation.

\begin{table}[H]
%%\begin{table*}[t]
	\caption{Performance of the baselines and our model on top-10 recommendation. 
		All the values in the table are percentage numbers with `\%' omitted, and all differences are significant at $p <0.05$.
		The stared numbers ($^*$) indicate the best baseline performances, and the bolded numbers indicate the best performance of each column. 
		The last line shows the percentage improvement of our model against the best baseline (i.e., JRL), which are significant at $p<0.001$.}
	\hspace{-0.9cm}
	\setlength{\tabcolsep}{0pt}
	\begin{tabular}
	{lllllllllllllllll} 
	\toprule
		\bf{Dataset} &  \multicolumn{4}{c}{\bf{CDs and Vinyl}} & \multicolumn{4}{c}{\bf{Clothing}} & \multicolumn{4}{c}{\bf{Cell Phones}} & \multicolumn{4}{c}{\bf{Beauty}}\\\hline
		\bf{Measures}(\%) & \bf{NDCG} &  \bf{Recall} &  \bf{HR} & \bf{Prec.} &\bf{ NDCG} & \bf{Recall} & \bf{HR} & \bf{Prec.} & \bf{NDCG} &\bf{ Recall} & \bf{HR} & \bf{Prec.} & \bf{NDCG} & \bf{Recall} & \bf{HR} & \bf{Prec.} \\\hline
		\bf{BPR}  & 2.009 & 2.679 & 8.554 & 1.085 & 0.601 & 1.046 & 1.767 & 0.185 & 1.998 & 3.258 & 5.273 & 0.595 & 2.753 & 4.241 & 8.241 & 1.143\\
		\bf{BPR-HFT} & 2.661 & 3.570 & 9.926 & 1.268 & 1.067 & 1.819 & 2.872 & 0.297 & 3.151 & 5.307 & 8.125 & 0.860 & 2.934  & 4.459 & 8.268 & 1.132\\
		\bf{VBPR} & 0.631 & 0.845 & 2.930 & 0.328 & 0.560 & 0.968 & 1.557 & 0.166 & 1.797 & 3.489 & 5.002 & 0.507 & 1.901 & 2.786 & 5.961 & 0.902\\
		\bf{TransRec} & 3.372 & 5.283 & 11.956 & 1.837 & 1.245 & 2.078 & 3.116 & 0.312 & 3.361 & 6.279 & 8.725 & 0.962 & 3.218  & 4.853 & 9.867 & 1.285\\
		\bf{DeepCoNN}  & 4.218 & 6.001 & 13.857 & 1.681 & 1.310 & 2.332 & 3.286 & 0.229 & 3.636 & 6.353 & 9.913 & 0.999 & 3.359 & 5.429 & 9.807 & 1.200\\
		\bf{CKE}  & 4.620 & 6.483 & 14.541 & 1.779 & 1.502 & 2.509 & 4.275 & 0.388 & 3.995 & 7.005 & 10.809 & 1.070 & 3.717 & 5.938 & 11.043 & 1.371\\
		\bf{JRL}%Please define the meaning of the * in below.
		 & {5.378} $^*$ & {7.545} $^*$ & {16.774} $^*$ & {2.085} $^*$ & {1.735} $^*$ & {2.989} $^*$ & {4.634} $^*$ & {0.442} $^*$ & {4.364} $^*$ & {7.510} $^*$ & {10.940} $^*$ & {1.096} $^*$ & {4.396} $^*$ & {6.949} $^*$ & {12.776} $^*$ & {1.546} $^*$\\\hline
		\bf{Our model} & \textbf{5.563} & \textbf{7.949} & \textbf{17.556} & \textbf{2.192} & \textbf{3.091} & \textbf{5.466} & \textbf{7.972} & \textbf{0.763} & \textbf{5.370} & \textbf{9.498} & \textbf{13.455} & \textbf{1.325} & \textbf{6.399} & \textbf{10.411} & \textbf{17.498} & \textbf{1.986}\\\hline
		\bf{Improvement} & 3.44 & 5.35  & 4.66 & 5.13 & 78.16 & 82.87 & 72.03 & 72.62 & 23.05 & 26.47 & 22.99 & 20.89 & 45.56 & 49.82 & 36.96 & 28.46\\
	\bottomrule
	\end{tabular}\label{tab:result}
%	\vspace{-5pt}
%%\end{table*}
\end{table}

%deep models performed better
%JRL perform best
%Our model is even better

%TODO: add parameter sensitivity analysis

%%\begin{figure}[H]
%%%\vspace{-15pt}
%%%\centering
%%%\hspace{-4pt}
%%	\begin{subfigure}[]{.4\textwidth}
%%		\centering
%%		\includegraphics*[viewport=7mm 7mm 195.5mm 145mm, scale=0.4]{./fig/cd-size-ndcg.pdf}%width=1.75in
%%		\label{fig:cd_size}
%%	\end{subfigure}%
%%	\hspace{40pt}
%%	\begin{subfigure}{.4\textwidth}
%%		\centering
%%		\includegraphics*[viewport=7mm 7mm 195.5mm 145mm, scale=0.4]{./fig/beauty-size-ndcg.pdf}%width=1.75in
%%		\label{fig:be_size}
%%	\end{subfigure}%
%%%	\vspace{-10pt}
%%	\caption{The NDCG@10 performance of our model (ECFKG, Explainable \hl{CF} over Knowledge Graph) and the baseline methods under different embedding sizes.}
%%	\label{fig:embed}
%%\end{figure}

\begin{figure}[H]
%\vspace{-15pt}
%\centering
%\hspace{-4pt}
	\begin{subfigure}[]{.4\textwidth}
		\centering
		\includegraphics*[viewport=7mm 7mm 195.5mm 145mm, scale=0.4]{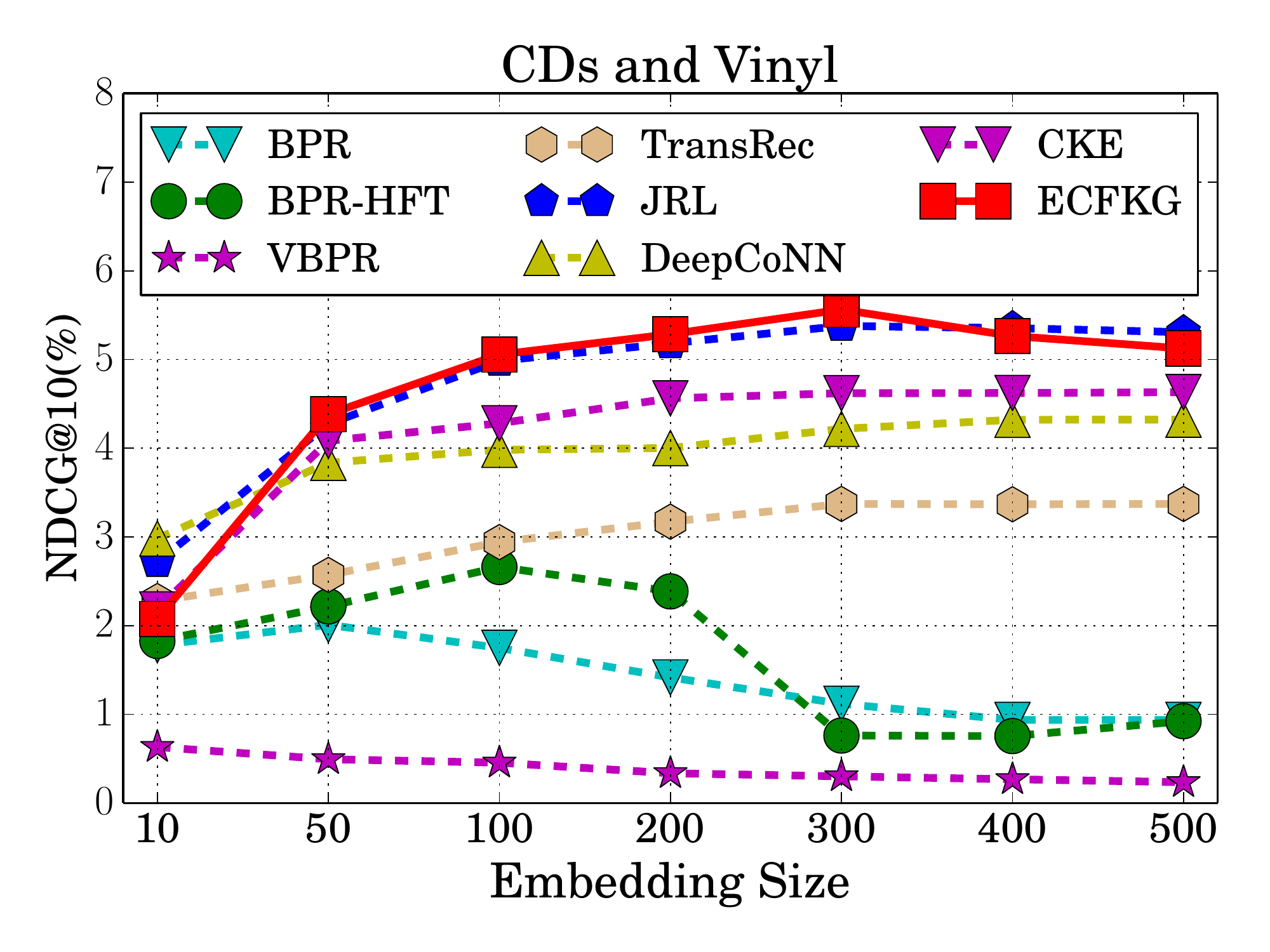}%width=1.75in
		\label{fig:cd_size}
	\end{subfigure}%
	\hspace{40pt}
	\begin{subfigure}{.4\textwidth}
		\centering
		\includegraphics*[viewport=7mm 7mm 195.5mm 145mm, scale=0.4]{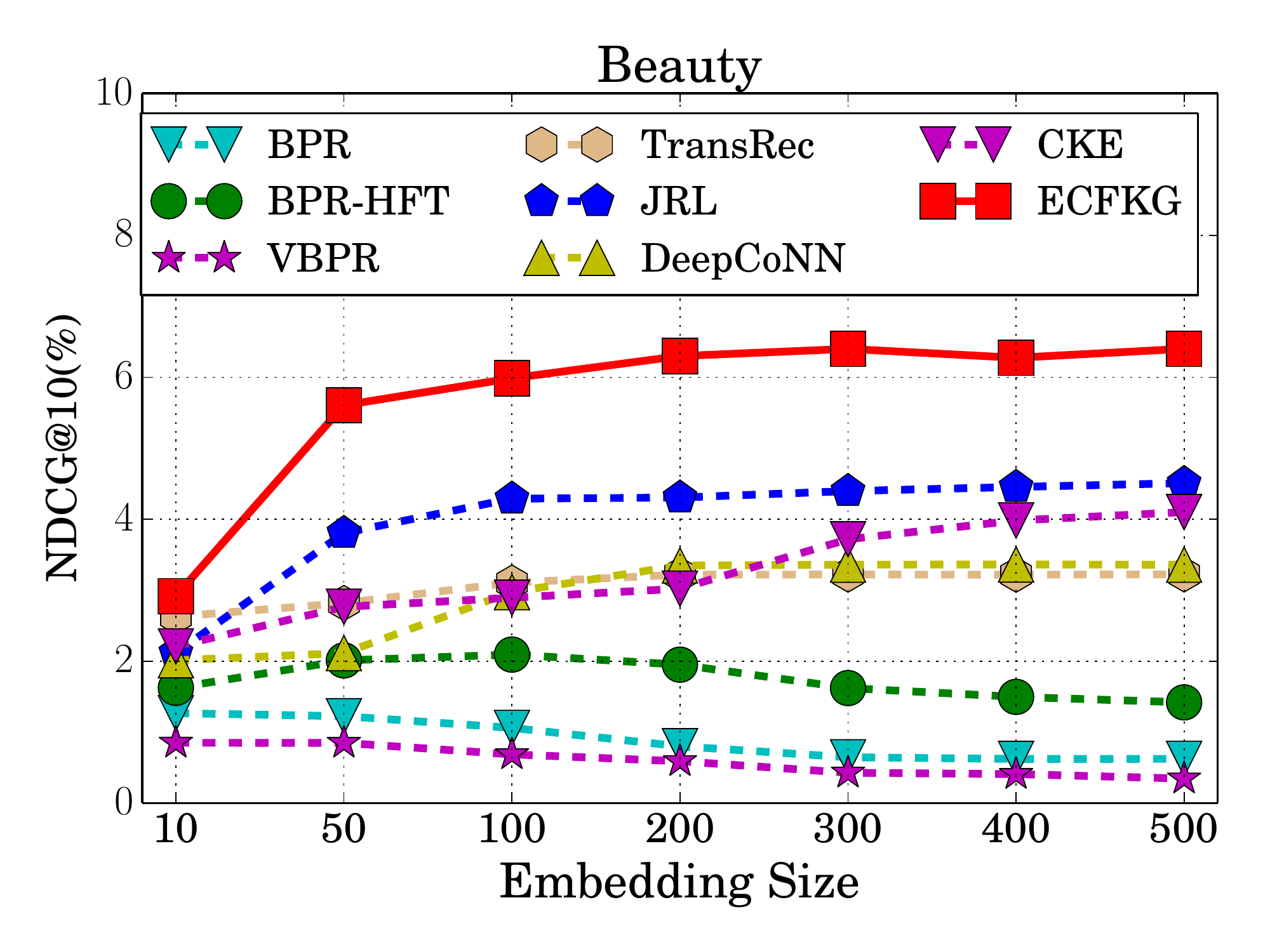}%width=1.75in
		\label{fig:be_size}
	\end{subfigure}%
%	\vspace{-10pt}
	\caption{The NDCG@10 performance of our model (ECFKG, Explainable Collaborative Filtering over Knowledge Graph) and the baseline methods under different embedding sizes.}
	\label{fig:embed}
\end{figure}

Figure~\ref{fig:embed} depicts the recommendation performance of our model and baseline methods with different embedding sizes on \textit{CDs \& Vinyl} and \textit{Beauty} datasets. Observations on the other two datasets are similar.
As shown in Figure~\ref{fig:embed}, the recommendation methods based on shallow models \linebreak{(BPR, BPR-HFT, and VBPR) }obtain the best NDCG when the embedding size is fairly small \linebreak{(from 10 to 100),} and larger embedding sizes usually hurt the performance of these models.
In~contrast to the shallow models, the results of neural models (i.e., JRL, DeepCoNN, CKE, and our model) show positive correlations with the increase of embedding sizes.
For example, the NDCG of the best baseline (JRL) and our model improves when the embedding size increases from 10 to 300, and remains stable afterwards. 
Overall, our model is robust to the variation of embedding sizes and consistently outperformed the baselines.

In Table~\ref{tab:result}, we see that the CKE model did not perform as well as we expected.
Although it has incorporated reviews, images and all other product knowledge described in this paper, the CKE model did not perform as well as JRL and our model.
One possible reason is that CKE only considers heterogeneous information in the construction of item representations, but it does not directly leverage the information for user modeling.
Another potential reason is that CKE separately constructs three latent spaces for text, image and other product knowledge, which makes it difficult for information from different types of data to propagate among the entities.
Either way, this indicates that the embedding-based relation modeling of our model is a better way to incorporate structured knowledge for product recommendation.

From the results we can also see that datasets of different density result in different performance in our model. In particular, denser datasets (\textit{Cell Phone} and \textit{Beauty}) generally get better ranking performance than sparser datasets (\textit{CD \& Vinyl} and \textit{Clothing}) in our model, \linebreak{which means} more sufficient information can help to learn better models in our algorithm.

To answer \textbf{RQ2}, we experiment the performance of our model when using different relations. Because we eventually need to provide item recommendations
for users, our approach would at least need the \textit{Purchase} relation to model the user purchase histories. As a result, we train and test our model built with only the \textit{Purchase} relation, as well as \textit{Purchase} plus one another relation separately.
%performance not equal to usefulness
As shown in Table~\ref{tab:info-result}, the relative performance of our models built on different relations varies considerably on the four datasets, which makes it difficult to conclude which type of product knowledge is the globally most useful one.
This, however, is not surprising because the value of relation data depends on the properties of the candidate products.
%In most cases, the incorporation of additional product knowledge boosts the performance of the recommender system.
On \textit{CDs and Vinyl}, where most products are music CDs, \linebreak{the CD} covers did not reveal much information, and people often express their tastes and preferences in the reviews they wrote. Thus \textit{Mention} turns out to be the most useful relation.
On \textit{Clothing}, however, reviews are not as important as the appearance or picture of the clothes, instead, it is easier to capture item similarities from the items that have been clicked and viewed by the same user.
Therefore, adding \textit{Also\_view} relation produces the largest performance improvement for our model on \textit{Clothing}.

\begin{table}[H]
%%\begin{table*}[t]
	\caption{Performance of our model on top-10 recommendation when incorporating \textit{Purchase} with other types of relation separately. 
		All the values in the table are percentage numbers with `\%' omitted, and all differences are significant at $p <0.05$. 
	}
	\hspace{-1cm}
	\setlength{\tabcolsep}{1pt}
	\begin{tabular}
		{lllllllllllllllll} \toprule
		\bf{Relations} &  \multicolumn{4}{c}{\bf{CDs and Vinyl}} & \multicolumn{4}{c}{\bf{Clothing}} & \multicolumn{4}{c}{\bf{Cell Phones}} & \multicolumn{4}{c}{\bf{Beauty}}\\\hline
		\bf{Measures}(\%) & \bf{NDCG} & \bf{Recall} & \bf{HT} & \bf{Prec} & \bf{NDCG} & \bf{Recall} & \bf{HT} & \bf{Prec} & \bf{NDCG} & \bf{Recall} & \bf{HT} & \bf{Prec} & \bf{NDCG} & \bf{Recall} & \bf{HT} & \bf{Prec} \\\hline
		\bf{\textit{Purchase} only} & 1.725&	2.319&	7.052&	0.818&	0.974&	1.665&	2.651&	0.254&	2.581&	4.526&	6.611&	0.649&	2.482&	3.834&	7.432&	0.948 \\\hline
		\bf{~+\textit{Also\_view}} & 1.722&	2.356&	6.967&	0.817&	1.800&	3.130&	4.672&	0.448&	2.555&	4.367&	6.417&	0.630&	4.592&	7.505&	12.901&	1.511\\
		\bf{~+\textit{Also\_bought}} & 3.641&	5.285&	12.332&	1.458&	1.352&	2.419&	3.580&	0.343&	4.095&	7.129&	10.051&	0.986&	4.301&	6.994&	11.908&	1.408\\
		\bf{~+\textit{Bought\_together}} & 1.962&	2.712&	7.473&	0.861&	0.694&	1.284&	2.026&	0.189&	3.173&	5.572&	7.952&	0.784&	3.341&	5.337&	9.556&	1.181\\
		\bf{~+\textit{Produced\_by}} & 1.719&	2.318&	6.842&	0.792&	0.579&	1.044&	1.630&	0.155&	2.852&	4.982&	7.274&	0.719&	3.707&	5.939&	10.660&	1.287\\
		\bf{~+\textit{Belongs\_to}} & 2.799&	4.028&	10.297&	1.200&	1.453&	2.570&	3.961&	0.376&	2.807&	4.892&	7.242&	0.717&	3.347&	5.382&	9.994&	1.193\\
		\bf{~+\textit{Mention}} & 3.822&	5.185&	12.828&	1.628&	1.019&	1.754&	2.780&	0.265&	3.387&	5.806&	8.548&	0.848&	3.658&	5.727&	10.549&	1.305\\ \hline
		\bf{~+\textit{all}} (our model) & \textbf{5.563} & \textbf{7.949} & \textbf{17.556} & \textbf{2.192} & \textbf{3.091} & \textbf{5.466} & \textbf{7.972} & \textbf{0.763} & \textbf{5.370} & \textbf{9.498} & \textbf{13.455} & \textbf{1.325} & \textbf{6.370} & \textbf{10.341} & \textbf{17.131} & \textbf{1.959}\\
		%Improvement & 3.44 & 5.35  & 4.66 & 5.13 & 78.16 & 82.87 & 72.03 & 72.62 & 23.05 & 26.47 & 22.99 & 20.89 & 44.90 & 48.81 & 34.09 & 26.71\\\hline
	\bottomrule
	\end{tabular}\label{tab:info-result}
%	\vspace{-5pt}
%%\end{table*}
%%\end{table*}
\end{table}

Overall, it is difficult to find a product knowledge that is universally useful for recommending products from all categories.
%However, we see that our final model, which integrates all the available knowledge relations, achieves the best performance on all four datasets, which 
%To build recommender systems in practice, we suggest testing different product knowledge in each product category separately and build models accordingly. 
However, we see that by modeling all of the heterogenous relation types, our final model outperforms all the baselines and outperforms all the simplified versions of our model with one or two types of relation, which implies that our KBE approach to recommendation is scalable to new relation types, and it has the ability to leverage very heterogeneous information sources in a unified manner.

To answer \textbf{RQ3}, we compare the training efficiency of different methods in our experiments on the same Nvidia Titan X GPU platform. The testing procedures for all methods are quite efficient, \linebreak{and generating} the recommendation list for a particular user on the largest \textit{CDs \& Vinyl} dataset requires less than 20 milliseconds. This is because after the model has learned the embeddings of the users and items, generating the recommendation list does not require re-computation of the embeddings and only needs to calculate their similarity. In terms of training efficiency, shallow models can be much more efficient than deep neural models, which is not surprising. For example, BPR or BPR-HFT can be trained within 30 minutes on the largest \textit{CDs \& Vinyl} dataset, while deep models such as DeepCoNN, CKE, and JRL takes about 10 hours on the same dataset, but they also bring much better recommendation performance. Our Explainable CF over Knowledge Graph (ECFKG) approach takes comparable training time on the largest dataset (about 10 hours), \linebreak{while achieving} better recommendation performance than other deep neural baselines, which is a good balance between efficiency and effectiveness.

\subsection{Case Study for Explanation Generation}

\begin{figure}[H]
	\includegraphics[width=6.5in]{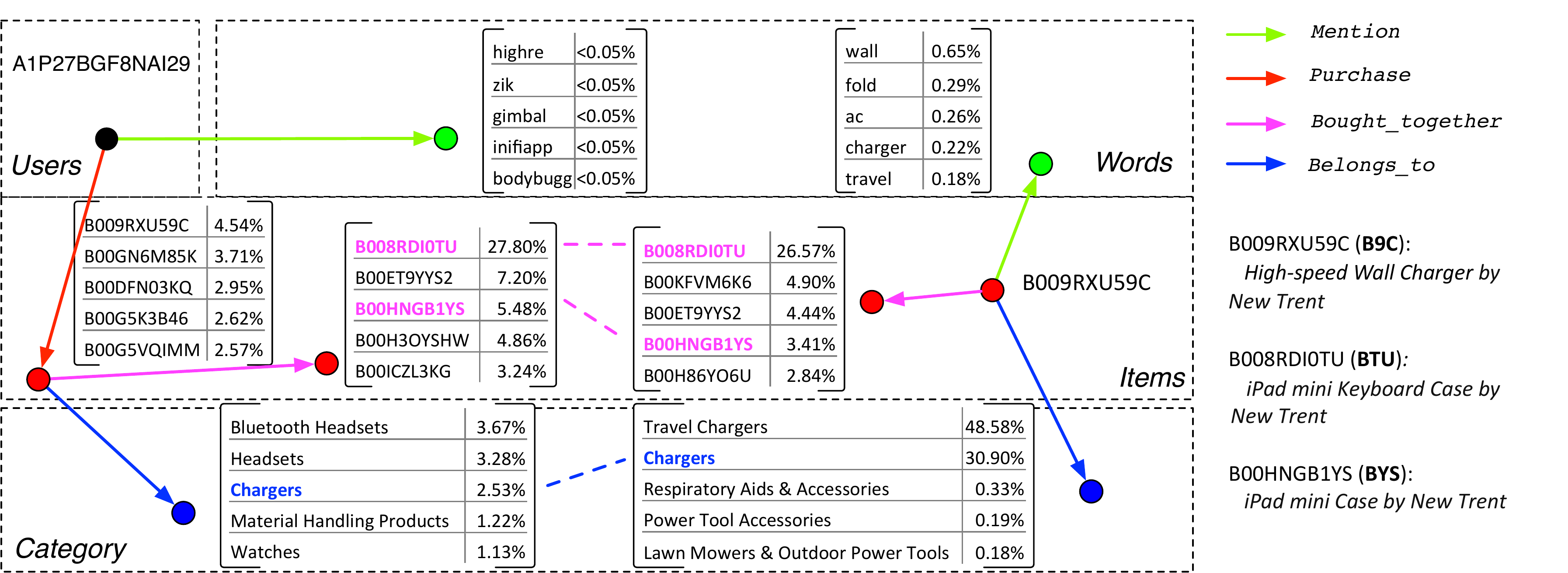}
	\caption{Example explanation paths between the user \textit{A1P27BGF8NAI29} and the item \textit{B009RXU59C} (B9C) in \textit{Cell Phones}.}%Please check and confirm.
	\label{fig:case_study}
%\end{figure*}
\end{figure}

To show the ability of our model to generate knowledge-enhanced explanations, we conduct case study for a test user (i.e., \textit{A1P27BGF8NAI29}) from \textit{Cell Phones}, for whom we have examined that the first recommendation (i.e., \textit{B009RXU59C}) provided by the system is correct.
We plot the translation process of this user to other entity subspaces with \textit{Purchase}, \textit{Mention}, \textit{Bought\_together}, and \textit{Belongs\_to} relations, as shown in Figure~\ref{fig:case_study}.
We also show the translation of the first recommended item \textit{B009RXU59C} (\textit{B9C}) using the same relations.
The top 5 entities retrieved by our system for each translation are listed along with their probabilities computed based on Equation~(\ref{equ:one_direction_prob}).

As we can see in Figure~\ref{fig:case_study}, there are three explanation paths between the user \textit{A1P27BGF8NAI29} and the item \textit{B9C}.
The first and second paths are constructed by \textit{Purchase} and \textit{Bought\_together}.
According to our model, the user is linked to \textit{B008RDI0TU} (\textit{BTU}) and \textit{B00HNGB1YS} (\textit{BYS}) through \textit{Purchase}$+$\textit{Bought\_together} with probabilities as 27.80\% and 5.48\%.
The item \textit{B9C} is linked to \textit{BTU} and \textit{BYS} through \textit{Bought\_together} directly with probabilities as 26.57\% and 3.41\%.
The third path is constructed by \textit{Purchase} and \textit{Belongs\_to}.
The user is linked to the category \textit{Chargers} with probability as 2.53\% and \textit{B9C} is linked to \textit{Chargers} with probability as 30.90\%.
Therefore, we can create three natural language explanations for the recommendation of \textit{B9C} by describing these explanation paths with simple templates.
Example sentences and the corresponding confidences are listed below:

$\bullet$ \textit{B9C} is recommended because the user often purchases items that are bought with \textit{BTU} together, and \textit{B9C} is also frequently bought with \textit{BTU} together ($27.80\%\times 26.57\%=7.39\%$).

\vspace{6 pt} $\bullet$ \textit{B9C} is recommended because the user often purchases items that are bought with \textit{BYS} together, and \textit{B9C} is also frequently bought with \textit{BYS} together ($5.48\%\times 3.41\%=0.19\%$).

$\bullet$ \textit{B9C} is recommended because the user often purchases items related to the category \textit{Chargers}, and \textit{B9C} belongs to the category \textit{Chargers} ($2.53\%\times 30.90\%=0.78\%$).

Among the explanation sentences, the best explanation should be the one with the highest confidence, which is the first sentence in this case.%%Is this a list or a paragraph? Please check and confirm.

To better evaluate the quality of these recommendation explanations, we look at the details of each product shown in Figure~\ref{fig:case_study}.
On Amazom.com, \textit{B9C} is a \textit{High-speed Wall Charger by New Trend} for tablets, \textit{BTU} is an \textit{iPad mini Keyboard Case}, and \textit{BYS} is an \textit{iPad Air Keyboard Case}.
If the generated explanations are reasonable, this means that the user has purchased some items that were frequently co-purchased with iPad accessories.
Also, this indicates that there is a high probability that the user has an iPad. 
For validation proposes, we list the five training reviews written by the user in Table~\ref{tab:user_review}.

As we can see in Table~\ref{tab:user_review}, the user has purchased several tablet accessories such as Bluetooth headsets and portable charger.  
The second review even explicitly mentions that the user has possessed an iPad and expresses concerns about the ``running-out-of-juice'' problem.
Therefore, it is reasonable to believe that the user is likely to purchase stuff that are frequently co-purchased with iPad accessories such as \textit{iPad mini Keyboard Case} (\textit{BTU}) or \textit{iPad Air Keyboard Case} (\textit{BYS}), which are recommended as top items in our algorithm, and are also well-explained by the explanation paths in the knowledge graph.
%explain the graph

\begin{table}[H]
	\centering
	\caption{The reviews written by the user \textit{A1P27BGF8NAI29} in the training data of \textit{Cell Phones}.}
	\begin{tabular}{p{15cm}}
%%		\hline\hline
           \toprule
		Review of Jabra VOX Corded Stereo Wired \textbf{Headsets} \\ 
%%		\hline
		$~~$\textit{... I like to listen to music at work, but I must wear some sort of \textbf{headset} so that I do not create a disturbance. So, I have a broad experience in \textbf{headsets} ...}\\ 
%%		\toprule
		\midrule
		Review of OXA Juice Mini M1 2600mAh \\
%%		 \hline
		$~~$\textit{... I recently had an experience, where I was about town and need to recharge my \textbf{iPad}, and so I tried this thing out. I plugged in the \textbf{iPad}, and it quickly charged it up, and at my next destination it was ready to go ...}\\ \hline
%%		\toprule
		Review of OXA 8000mAh Solar External Battery Pack Portable\\ 
%%		\hline
		$~~$\textit{... This amazing gadget is a solar powered \textbf{charger} for your small electronic device. This \textbf{charger} is (according to my ruler) ~5-1/4 inches by ~3 inches by about 6 inches tall. So, it is a bit big to place in the pocket ...}\\\hline 
%%		\toprule
		Review of OXA Bluetooth Wristwatch Bracelet \\ 
%%		\hline
		$~~$\textit{... I was far from thrilled with \textbf{Bluetooth headset} that I had, so I decided to give this device a try. Pros: The bracelet is not bad looking, ...}\\ 
%%		\toprule
		Review of OXA Mini Portable Wireless \textbf{Bluetooth} Speaker \\ 
%%		\hline
		$~~$\textit{... This little gadget is a \textbf{Bluetooth} speaker. It's fantastic! This speaker fits comfortably in the palm of your hand, ...}\\ 
%%		\hline\hline
      \bottomrule
	\end{tabular}
	\label{tab:user_review}
\end{table}

\section{Conclusions and Outlook}\label{sec:conclusions}
%In this paper, we proposed to integrate user multi-type behaviors into a unified graph, based on which we extended traditional collaborative filtering based on this graph for more accurate user profiling and recommendations. Experimental results demonstrate the effectiveness of our designed model.

In this paper, we propose to learn over heterogenous KBE for personalized explainable recommendation. To do so, we construct the user-item knowledge graph to incorporate both user behaviors and our knowledge about the items. We further learn the KBE with the heterogenous relations collectively, and leverage the user and item embeddings to generate personalized recommendations. To explain the recommendations, we devise a soft matching algorithm to find explanation paths between a user and the recommended items in the latent KBE space.
%and each explanation path can be described as a natural language sentence for recommendation explanation. 
Experimental results on real-world datasets verified the superior performance of our approach, as well as its flexibility to incorporate multiple relation types.

After years of success at integrating machine learning into recommender systems, we believe that equipping the systems with knowledge (again) is important to the future of recommender \linebreak{systems -- or }information systems in a broader sense --- which can help to improve both the performance and explainability in the future.

%There are many directions to extend this work. For example, to make our recommendation more transparent, we can conduct reasoning over the learned user behavior graph to provide explanations for the recommended items. Then, we can design more advanced transfer function to power our model for capturing more accurate properties.


%=====================================
% References, variant A: internal bibliography
%=====================================
\iffalse

\reftitle{References}

\fi%
% The following MDPI journals use author-date citation: Arts, Econometrics, Economies, Genealogy, Humanities, IJFS, JRFM, Laws, Religions, Risks, Social Sciences. For those journals, please follow the formatting guidelines on http://www.mdpi.com/authors/references
% To cite two works by the same author: \citeauthor{ref-journal-1a} (\citeyear{ref-journal-1a}, \citeyear{ref-journal-1b}). This produces: Whittaker (1967, 1975)
% To cite two works by the same author with specific pages: \citeauthor{ref-journal-3a} (\citeyear{ref-journal-3a}, p. 328; \citeyear{ref-journal-3b}, p.475). This produces: Wong (1999, p. 328; 2000, p. 475)

%=====================================
% References, variant B: external bibliography
%=====================================
\reftitle{References}

\end{document}